\documentclass[letterpaper, 10 pt, conference]{ieeeconf} 
\pdfoutput=1

\IEEEoverridecommandlockouts                              
\usepackage{eso-pic}
\usepackage{graphics} 
\usepackage{epsfig} 
\usepackage{mathptmx} 
\usepackage{times} 
\usepackage{amsmath} 
\usepackage{amssymb}  
\usepackage[utf8x]{inputenc}
\usepackage[T1]{fontenc}
\usepackage{listings}
\usepackage{lscape}
\usepackage{subfigure}
\usepackage{amsfonts}
\usepackage{graphicx}
\usepackage{hyperref}
\hypersetup{colorlinks=true,linkcolor=black,filecolor=magenta,urlcolor=cyan}

\title{\LARGE \bf
Time-optimal control of cranes subject to container height constraints}

\author{Filipe Marques Barbosa and Johan Löfberg
\thanks{This work was supported by VINNOVA Competence Center LINK-SIC.}
\thanks{F. Marques Barbosa and J. Löfberg are with the Division of Automatic Control, Department of Electrical Engineering, Linköping University, Linköping, Sweden
{\tt\small filipe.barbosa@liu.se, johan.lofberg@liu.se}}%
\thanks{\small DOI: \href{https://doi.org/10.23919/ACC53348.2022.9867816}{10.23919/ACC53348.2022.9867816}}
}

\begin{document}

\AddToShipoutPictureBG*{%
\put(0,20){
\hspace*{\dimexpr0.075\paperwidth\relax}
\parbox{.84\paperwidth}{\footnotesize \copyright2022 IEEE. Personal use of this material is permitted.  Permission from IEEE must be obtained for all other uses, in any current or future media, including reprinting/republishing this material for advertising or promotional purposes, creating new collective works, for resale or redistribution to servers or lists, or reuse of any copyrighted component of this work in other works.}%
}}

\maketitle
\thispagestyle{empty}
\pagestyle{empty}

\begin{abstract}
	
The productivity and efficiency of port operations strongly depend on how fast a ship can be unloaded and loaded again. With this in mind, ship-to-shore cranes perform the critical task of transporting containers into and onto a ship and must do so as fast as possible. Though the problem of minimizing the time spent in moving the payload has been addressed in previous studies, the different heights of the container stacks have not been the focus. In this paper, we perform a change of variable and reformulate the optimization problem to deal with the constraints on the stack heights. As consequence, these constraints become trivial and easy to represent since they turn into bound constraints when the problem is discretized for the numerical solver. To validate the idea, we simulate a small-scale scenario where different stack heights are used. The results confirm our idea and the representation of the stack constraints become indeed trivial. This approach is promising to be applied in real crane operations and has the potential to enhance their automation.

\end{abstract}

\section{INTRODUCTION}

Shipping containers revolutionized the transportation of goods across the globe. Before its introduction, the process of loading and unloading a ship was extremely slow and they had to stay for weeks in the port at a time. Thanks to containerization, the productivity and efficiency of ports have been remarkably enhanced. Moreover, an important performance indicator of a port is the speed at which a ship is unloaded and loaded again \cite{Kreuzer2014,Sakawa1982}.

\begin{figure}[h]
	\centering
	\includegraphics[scale=0.17]{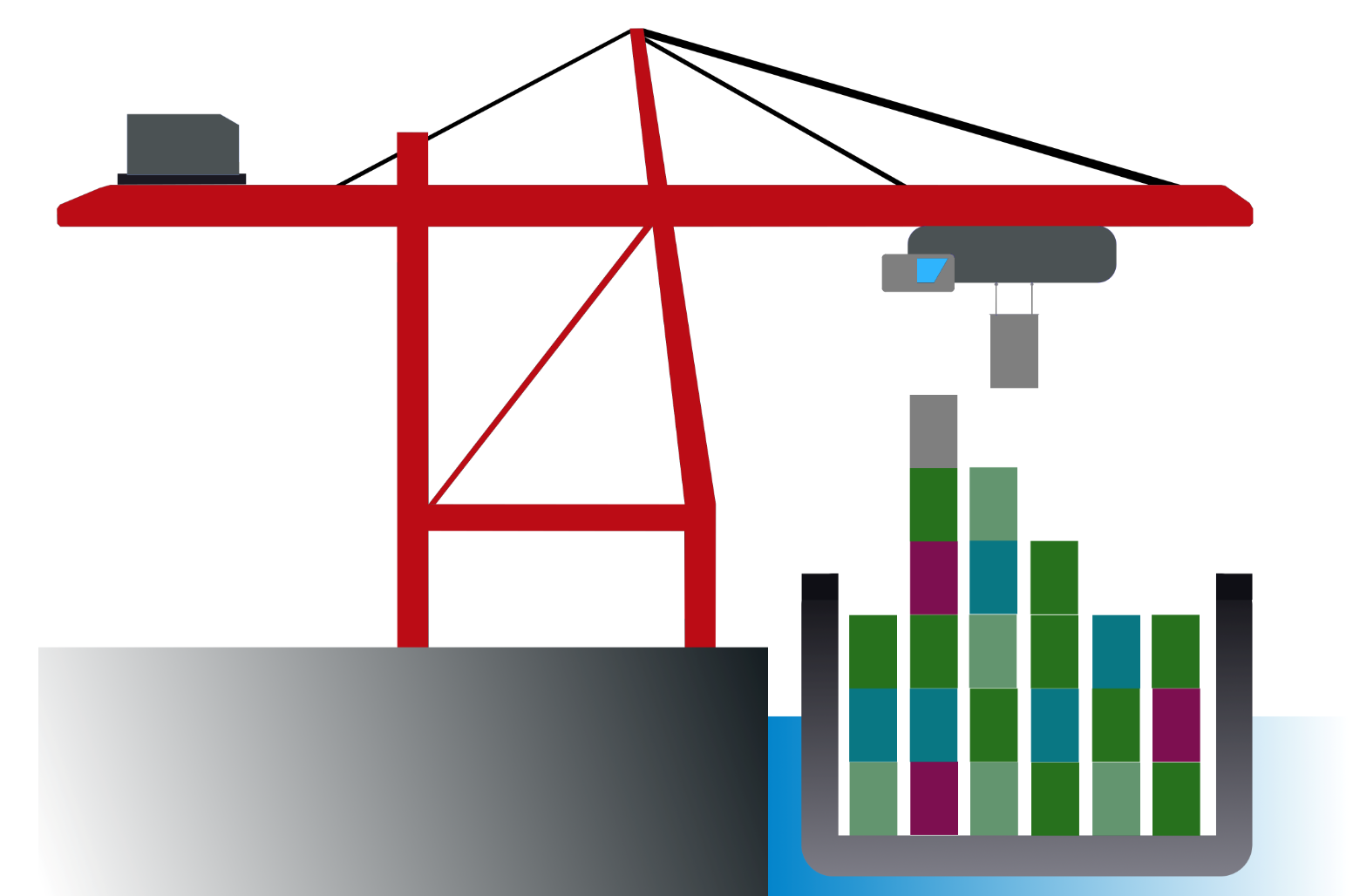}
	\caption{A ship-to-shore crane schematic.}
	\label{fig:sts}
\end{figure}

With this in mind, the efficiency of port operations deeply depends on the performance of the ship-to-shore (STS) cranes. They are responsible for the critical task of loading and safely unloading the ship \cite{Kreuzer2014,Sakawa1982,Arena2015}. See \autoref{fig:sts}. Thus making their operations faster is crucial for the performance indicators of the port.

Because of this, various approaches have been proposed to address the faster loading and unloading of containers. Some authors have focused on minimizing the swing and other oscillations to minimize the time of the loading and unloading process. For instance, in the works done by Kreuzer \textit{et. al.,} \cite{Kreuzer2014} and Kim \textit{et. al.,} \cite{kim2004anti}, where the main idea is that a good estimation of the sway angle will make it easier to be mitigated and thus lead to faster operations. In the former, an unscented Kalman filter is designed to provide an accurate estimation of the payload swing. Then, it is validated by applying it in different control techniques. In the latter, an anti-sway controller is designed without the use of a vision system, thus making its implementation cheaper. 

Following this, optimal control has also been used in crane operations as a combination of both anti-sway and trolley motion control. Sakawa and Shindo \cite{Sakawa1982} divide the payload motion into five sections and derive an optimal speed reference trajectory, and the swing of the payload is then minimized. Furthermore, Maghsoudi \textit{et al.,} \cite{Maghsoudi2016} have used a three-dimensional model of the crane and tuned a PID controller by using a cost function with the aim of fast payload positioning and minimum sway.

Moreover, time-optimal control has been used as a natural approach when one aims to make the loading and unloading as fast as possible. In such approaches, the objective is to minimize the final time $t_f$. That is, the total time for the payload to go from an initial position at rest to a desired final position at rest again.

With this in mind, Auernig and Troger \cite{Auernig1987} use the Pontryagin maximum principle and obtain the minimum-time solution for moving the payload. Also, based on differential flatness Chen \textit{et al.,} \cite{Chen2016} propose a time-optimal offline trajectory planning method and a tracking controller.

Furthermore, time-optimal control has also been combined with anti-sway. Al-Garni \textit{et al.,} \cite{AlGarni1995} use a nonlinear dynamic model and constructs a cost function according to a performance index. This performance index is measured under the condition of making the process of loading or unloading in a minimum time as well as reducing the payload sway. Da Cruz and Leonardi \cite{DaCruz2012} use linear programming to solve the minimum-time anti-sway motion problem. The solution is obtained by solving a sequence of fixed-time maximum-range problems.

Those all are, without doubt, valuable contributions to the field of control of overhead cranes. However, a problem with much of the literature is that they do not address the different heights that the container stacks can have along the path from the shore to a specific position on the ship. As mentioned before, some works focus on estimating the sway in order to reduce it and consequently the transferring time e.g., \cite{Kreuzer2014} and \cite{kim2004anti}. Thus, hoisting movements are naturally not addressed. Furthermore, some works that explicitly deal with minimum time do not address hoisting at all. A constant length of the rope is used in \cite{Maghsoudi2016} and \cite{Chen2016}, and it is considered given in \cite{DaCruz2012}. Moreover, works dealing with hoisting do not take stack heights into account e.g., \cite{Sakawa1982}, \cite{Auernig1987}  and \cite{AlGarni1995}.

In this paper, we propose a time-optimal control approach to address the problem of loading and unloading a ship in the minimum time. Additionally, the different and arbitrary heights that the container stacks can have are taken into account here. Moreover, the novelty of this work is an easy and intuitive way to represent the stack heights into the optimization problem. This is made possible through a change of variable and reformulation of the optimal control problem. After the variable change, time $t$ becomes a state variable and the payload coordinate along the horizontal axis $x_p$ becomes the free variable. Now with the payload position as the free variable instead of time, the container avoidance constraints will turn into bound constraints when the problem is discretized for the numerical solver. This means that no functional representation of the stack heights is required.

This paper is organized as follows: \autoref{sec:modelling_and_original} presents the nonlinear model used to describe the crane movements and the original problem formulation, with time as the free variable. Which has been the standard way to solve this problem. \autoref{sec:reformulation} brings the variable change and the optimal control problem reformulation. \autoref{sec:geometric-constraints} shows how the stack heights are now represented along the horizontal axis, leading to a form that is suitable for numerical optimization. \autoref{sec:results} depicts an example to illustrate the proposed idea and subsequently discuss it. Lastly, \autoref{sec:conclusion} brings the conclusion of the work.

\section{MODELLING AND ORIGINAL PROBLEM FORMULATION}\label{sec:modelling_and_original}

In order to control an STS-crane and make the process of loading and unloading the ship as fast as possible, a model description of its dynamics is obtained and a time-optimal control problem solved. In this section, we first present a lumped mass model, following the single-rope hoisting mechanism. Subsequently, the standard formulation for time-optimal control of this system is presented.

\subsection{Modelling}\label{sec:modelling}

\begin{figure}[h]
	\centering
	\includegraphics[scale=0.6]{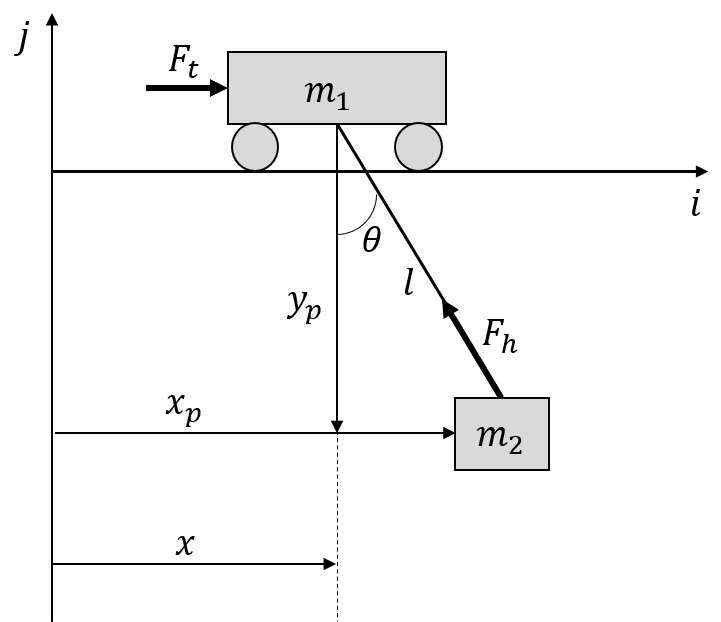}
	\caption{Two-dimensional schematic representing the trolley and payload motions.}
	\label{fig:schematic}
\end{figure}

A simplified representation of the dynamics of overhead cranes can be made by considering the problem as cart-pendulum with hoisting. To this end, consider the position of the trolley $x(t)$, the sway $\theta(t)$, the length of the hoisting rope $l(t)$ and, the payload coordinates $x_p(t)$ and $y_p(t)$ as generalized coordinates. The forces $F_t(t)$ and $F_h(t)$ are the control inputs applied to the trolley and for hoisting the payload, respectively. A two-dimensional schematic of this representation showing the general coordinates and the forces involved in the problem is depicted in \autoref{fig:schematic}.

Similarly to the model obtained in \cite{kim2004anti} and \cite{Hong2019}, and additionally including the position of the payload in general coordinates, the following equations of motion are obtained
\begin{multline}
	\label{eq:dynamics-bridge}
	(m_1+m_2)\ddot{x}(t) + m_2l(t)(\ddot{\theta}(t)\cos(\theta(t))-\dot{\theta}^2(t)\sin(\theta(t))) +\\
	m_2\ddot{l}(t)\sin(\theta(t))+2m_2\dot{l}(t)\dot{\theta}(t)\cos(\theta(t)) = F_t(t)
\end{multline}
\begin{multline}
	\label{eq:dynamics-hoist}
	m_2\ddot{l}(t) - m_2l(t)\dot{\theta}^2(t) - m_2g\cos(\theta(t)) +\\ m_2\ddot{x}(t)\sin(\theta(t)) = -F_h(t)
\end{multline}
\begin{equation}
	\label{eq:dynamics-sway}
	l(t)\ddot{\theta}(t) + 2\dot{l}(t)\dot{\theta}(t)+g\sin(\theta(t)) + \ddot{x}(t)\cos(\theta(t)) = 0
\end{equation}
\begin{equation}
	\label{eq:dynamics-xp}
	x_p(t) = \sin(\theta(t))l(t) + x(t)
\end{equation}
\begin{equation}
	\label{eq:dynamics-yp}
	y_p(t) = \cos(\theta(t))l(t).
\end{equation}
Here $g$ is the acceleration of gravity, and $m_1$ and $m_2$  are the trolley and payload masses, respectively. The dynamics of the bridge along the $i\text{-axis}$ is represented in (\ref{eq:dynamics-bridge}), the hoist motion of the payload in (\ref{eq:dynamics-hoist}), the sway motion of the payload in (\ref{eq:dynamics-sway}), and finally the payload position along $i\text{ and }j\text{-axis}$ are represented in (\ref{eq:dynamics-xp}) and (\ref{eq:dynamics-yp}), respectively. Note that the dynamics of the actuators themselves are not considered, and this model is formulated considering the dynamics of an overhead crane with a single hoisting rope mechanism. Moreover, the payload is treated as a point mass. 

Now, choosing the state variables as $x_1 = x_p$, $x_2 = \dot{x}_p$, $x_3 = y_p$, $x_4 = \dot{y}_p$, $x_5 = l$, $x_6 = \dot{l}$, $x_7 = \theta$ and $x_8 = \dot{\theta}$, the nonlinear state-space representation is derived and can be written as:
\begin{equation}
\label{eq:original-state-eq}
\begin{aligned}
		\dot{x}_{1} &= x_{2}\\
		\dot{x}_{2} &= -(u_{2}\sin(x_{7}))/m_{2} \\
		\dot{x}_{3} &= x_{4}\\
		\dot{x}_{4} &= -(u_{2}\cos(x_{7}))/m_{2}+g\\
		\dot{x}_{5} &= x_{6}\\
		\dot{x}_{6} &= x_{5}x_{8}^{2}+g\cos(x_{7})-u_{2}/m_{2}-\sin(x_{7})(u_{1}+u_{2}\sin(x_{7}))/m_{1}\\
		\dot{x}_{7} &= x_{8}\\
		\dot{x}_{8} &= -(2x_{6}x_{8}+g\sin(x_{7})+\cos(x_{7})(u_{1}+u_{2}\sin(x_{7}))/m_{1})/x_{5}\\
\end{aligned}
\end{equation}
with the control inputs $u_{1} = F_{t}$ and $u_{2}=F_{h}$.

With the equations of the motion of the crane written in the state-space form, the time-optimal control problem can be formulated.

\subsection{Original problem formulation}\label{sec:orignal_prob}

The problem of moving the payload from the initial to the final position in the minimum time can be expressed as a time-optimal control problem. This way, we need to minimize a cost function that accounts for the total time spent transporting the payload. This is done by obtaining a control law that satisfies the lower and upper bounds of the decision variables and control inputs.
\begin{equation}
\label{eq:original-formulation}
\begin{aligned}
& \underset{}{\text{minimize}}
& & T = \int_{0}^{t_f} \, dt \\
& \text{subject to}
& & \dot{x}(t) = f(t,x(t),u(t))\\
&&& x(0) = x_{0}\\
&&& x(t_f) = x_{t_f}\\
&&& g(x(t)) \le x(t) \le f(x(t))\\
&&& u_{min}(t) \le u(t) \le u_{max}(t).
\end{aligned}
\end{equation}

However, as mentioned before, STS-crane operations are usually subject to geometric constraints, corresponding to the height of the container stacks. Such constraints are usually nonlinear and non-smooth functions of space. Because of this, issues arise in formulating and solving the numerical optimization problem starting from (\ref{eq:original-formulation}).

The most immediate comes from our necessity of constructing the constraints on the container heights. This means that the geometric constraints in the numerical model need to be constructed from a function $g(x_p(t))$. Due to the discretization of the problem, the required height that the payload needs to be at a certain time instance $t^{k}$ is a function $g(x_p(t^{k}))$. Please note that in this paper we use superscript $k$ to indicate discretization. In this way, the function $g(x_p(t))$ needs to be defined from the current configuration using either continuous approximation or exact integer programming representation. Moreover, any nontrivial configuration of container heights will lead to non-convex constraints.

Furthermore, though the objective function looks simple, it cannot be directly used in a numerical optimization scheme. Once again, the problem here comes due to the discretization. Since the free variable is also the one being optimized, the solution will be influenced by the fixed time sampling rate, or fixed number of control intervals. Thus, it requires a strategy to be used, for instance, time-elastic bands, presented in \cite{Rosmann2015}.

In order to circumvent these problems, we reformulate the optimization problem through a change of variable.

\section{PROBLEM REFORMULATION}\label{sec:reformulation}

As mentioned in the previous section, container avoidance constraints are problematic when the independent variable is time $t$. That is, the stack heights are difficult to be represented when using time as the variable to be discretized. However, they can be easily described in the position along the $i\text{-axis}$.

With this in mind, the optimization problem is reformulated in an attempt to alleviate the above-mentioned issues. This is done by reparametrizing it in a spatial coordinate. Thus, we redefine the optimization problem with the position of the payload along the $i\text{-axis}$ $x_{p}$ as the variable that we discretize the dynamics along with. In other words, this means that we perform a change of variable and now the system's dynamics are described using $dx/dx_{p}$.

\subsection{The variable change}\label{sec:variable-change}

A change of the integration variable is now performed and the time-optimal problem is reformulated in the same fashion as in \cite{Verscheure2009} and \cite{KangShin1985}. First, consider the total time $T$ that the payload takes to go from the initial to the final position i.e., the cost function in (\ref{eq:original-formulation}) is
\begin{equation}
T = \int_{0}^{t_f} dt.
\end{equation}
Now, note that the first dynamic equation, used to describe the payload position along the $i\text{-axis}$, in (\ref{eq:original-state-eq}) is 
\begin{equation}
\label{eq:change_state}
\dot{x}_1=\frac{dx_1}{dt} = x_2 \implies \frac{dt}{dx_1} = \frac{1}{x_2}.
\end{equation}
In this way, the cost function to be minimized is rewritten as 
\begin{equation}
	\label{eq:obj-reformulation}
	T = \int_{0}^{t_f} dt = \int_{x_{1_0}}^{x_{1_f}} \frac{dx_1}{x_2} = \int_{x_{p_0}}^{x_{p_f}} \frac{dx_p}{x_2}.
\end{equation}

The variable change implies that the optimization problem is now parametrized in the payload position along the $i\text{-axis}$ ($x_{1}=x_{p}$) and consists of minimizing the integral on the right-hand side of (\ref{eq:obj-reformulation}). The main benefit of this will be discussed in the next section. Nevertheless, an immediate consequence of (\ref{eq:obj-reformulation}) is that a new state vector $x=[t,\dot{x}_p,y_p,\dot{y}_p,l,\dot{l},\theta, \dot{\theta}]^T$ is defined for the problem.

\textit{Remark:} With the variable change, time $t$ is no longer the free variable. Now, all the derivatives will be with respect to $x_p$ and thus, with different notation, $x^\prime=dx/dx_p$.

Additionally, since time $t$ is now a state variable, we make the following identification
\begin{equation}
	\label{eq:identifications}
	\setlength{\arraycolsep}{1pt}
	x_{1}\leftarrow{t},~x^\prime_{j}\leftarrow{\frac{dx_{j}}{dx_{p}}},~ j = 1, \ldots, n,
\end{equation}
and the state equations in (\ref{eq:original-state-eq}) become a system of differential algebraic equations
\begin{equation}
\label{eq:reformulated-state-eq}
\begin{aligned}
x_{2}x^\prime_{1} &= 1\\
x_{2}x^\prime_{2} &= -(u_{2}\sin(x_{7}))/m_{2}\\
x_{2}x^\prime_{3} &= x_{4}\\
x_{2}x^\prime_{4} &= -(u_{2}\cos(x_{7}))/m_{2}+g\\
x_{2}x^\prime_{5} &= x_{6}\\
x_{2}x^\prime_{6} &= \resizebox{0.85\hsize}{!}{$x_{5}x_{8}^{2}+g\cos(x_{7})-u_{2}/m_{2}-\sin(x_{7})(u_{1}+u_{2}\sin(x_{7}))/m_{1}$}\\ 
x_{2}x^\prime_{7} &= x_{8}\\
x_{2}x^\prime_{8} &= \resizebox{0.85\hsize}{!}{$ -(2x_{6}x_{8}+g\sin(x_{7})+\cos(x_{7})(u_{1}+u_{2}\sin(x_{7}))/m_{1})/x_{5},$} 
\end{aligned}
\end{equation}
where $j$ indices the state variables in the new state vector and $n$ is the system's dimension. It is worth noting that in performing the coordinate change, we implicitly make an assumption that the payload is moving monotonically, in one direction along $i\text{-axis}$. This is a natural limitation that implicitly implies a no-sway condition in the solution.

Subsequently to the change of variable made in (\ref{eq:obj-reformulation}), a natural choice of the cost function $J$ would be
\begin{equation}
	J = \int_{x_{p_{0}}}^{x_{p_{f}}} \frac{1}{x_2(x_p)} dx_p,
\end{equation}
which leads to convergence issues since $x_{2}(x_{p})=0$ at $x_{2}(0)$ and $x_{2}(x_{2_f})$. However, an interesting property of the reformulation is that the variable $x_1$ in (\ref{eq:reformulated-state-eq}) corresponds to time $x_{1}(x_{p})=t$. Thus, the optimization problem can now be written as 
\begin{equation}
\label{eq:opt-reformulated}
\begin{aligned}
& \underset{}{\text{minimize}}
& & J = t(x_{p_{f}}) \\
& \text{subject to}
& & x_{2}x^\prime(x_p) = f(x_p,x(x_p),u(x_p))\\
&&& x(0) = x_{0}\\
&&& x(x_{p_{f}}) = x_{f}\\
&&& 0 \le t(x_p)\\
&&& 0 \le x_2(x_p)\\
&&& g(x_p) \le x(x_p) \le f(x_p)\\
&&& u_{min}(x_p) \le u(x_p) \le u_{max}(x_p).
\end{aligned}
\end{equation}
As mentioned before and more evident now, the positions, velocities and accelerations are related to one another through the parametrization of path. This is an interesting feature for representing the geometric constraints as we will see in the next section. Furthermore, though the general dynamics remain nonlinear and non-convex, the cost function is nevertheless still convex, which is an additional outcome of the reformulation.

\section{GEOMETRIC CONSTRAINTS}\label{sec:geometric-constraints}

When dealing with the constraints that the different heights of the stacks impose on the payload height $y_p(x_p)$, a function $s(x_p)$ that represents the stack profile along the loading site is implicitly required. This way, the constraints on $y_p(x_p)$ are
\begin{equation}
0 \le y_{p}(x_{p}) \le h-s(x_{p}),
\label{eq:height_constraint_general}
\end{equation}
where $h$ is the maximum height, e.g., the distance from the ground to the trolley.

With the optimization problem in the original formulation as in (\ref{eq:original-formulation}), after the time discretization,  the constraints imposed on $y_p(x_p)$ in (\ref{eq:height_constraint_general}) would be
\begin{equation}
0 \le y_{p}(x_{p}(t^k)) \le h-s(x_{p}(t^k)).
\label{eq:height_constraint_time}
\end{equation}
Moreover, the required explicit function representation $s(x_{p}(t^k))$ will generally be discontinuous, nonlinear and non-convex. 

\begin{figure}[h]
	\centering
	\includegraphics[scale=0.39]{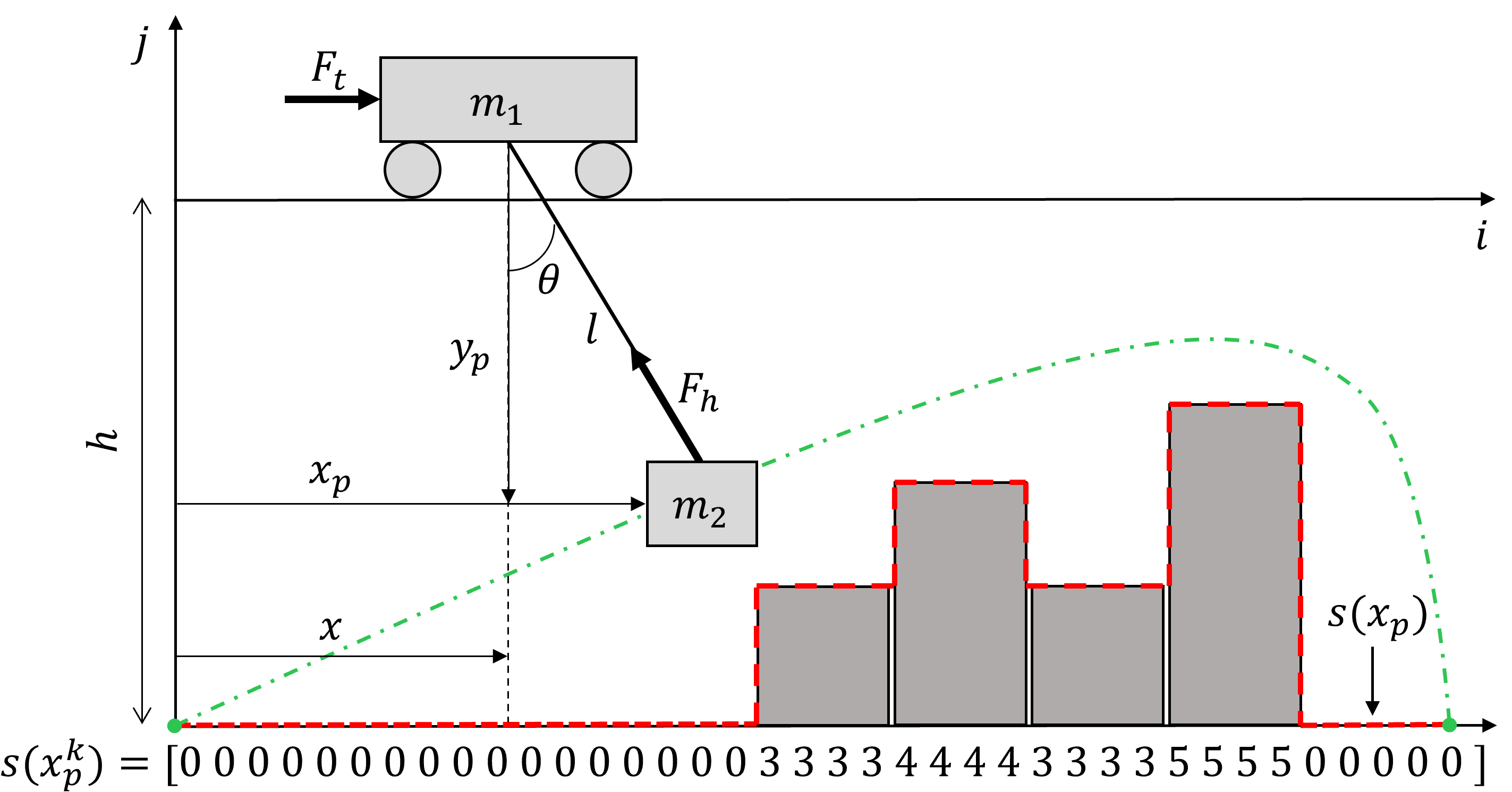}
	\caption{Two-dimensional schematic representing the trolley and payload motions and the geometric constraints (stacks).}
	\label{fig:schematic-constraint}
\end{figure}

However, with the optimization problem (\ref{eq:opt-reformulated}) now parametrized in $x_{p}$, the geometric constraints can be easily represented. The height of the stacks and their positions along the trajectory are represented by the function $s(x_{p})$, which now addresses the stack height at each position $x_{p}$. This way, when discretizing $x_{p}$, the function $s(x_{p})$ determines the upper bound constraints for $y_{p}(x_{p})$ in (\ref{eq:opt-reformulated}) as
\begin{equation}
	0 \le y_{p}(x_{p}^{k}) \le h-s(x_{p}^{k}).
	\label{eq:height_constraint_space}
\end{equation}
Note that we no longer need an explicit function $s(x_p)$, but simply function values that can be computed when setting up the numerical model. See \autoref{fig:schematic-constraint}.

This way of representing the stack heights and their positions is attractive since it is easier and more intuitive to be constructed and manipulated, after the discretization. This is the main benefit of the variable change in (\ref{eq:change_state})-(\ref{eq:obj-reformulation}) and constitutes the main contribution of this paper.

\section{RESULTS AND DISCUSSION}\label{sec:results}

In order to illustrate and validate the idea, a scenario of stack configuration was simulated. For simplicity, a small-scale example was used, where the integration goes from position $x_{p_{0}} = 0$ to $x_{p_{f}} = 1$. In this scenario, the container stacks are particularly high at the end of the loading site.

The software used to obtain the results was MATLAB. Additionally, CasADi software tool \cite{Andersson2019} and Yop toolbox \cite{leek2016optimal} were used for modeling the optimization problem, and IPOPT \cite{Wchter2005} was used to solve it.

Now, it is important to note that all the distances and lengths used in the results are in \textit{meters} and the angles in \textit{radian}. With (\ref{eq:opt-reformulated}) in mind, the initial and final conditions were set to
\begin{equation}
\label{eq:initial-final}
\begin{aligned}
&t(0) = 0,~ 				&&\\
&x_{p}(0) = 0, 	    		&&x_{p}(x_{p_{f}}) = 1,\\
&\dot{x}_p(0) = 0,~			&&\dot{x}_p(x_{p_f}) = 0,\\
&y_p(0) = 3,~				&&y_p(x_{p_f}) = 3,\\
&\dot{y}_p(0) = 0,~			&&\dot{y}_p(x_{p_f}) = 0,\\
&l(0) = 3,~					&&l(x_{p_f}) = 3,\\
&\dot{l}(0) = 0,~			&&\dot{l}(x_{p_f}) = 0,\\
&\theta(0) = 0,~			&&\theta(x_{p_f}) = 0,\\
&\dot{\theta}(0) = 0,~		&&\dot{\theta}(x_{p_f}) = 0,
\end{aligned}
\end{equation}
the constraints regarding the minimum and maximum height that the payload can be, given its position $x_p$ along the $i\text{-axis}$, were defined as
\begin{equation}
0.15 \le y_p(x_p) \le h-s(x_p),
\end{equation}
and the other box constraints were
\begin{equation}
\begin{aligned}
	0 \le &t(x_p) \\
	0 \le &\dot{x}_p(x_p)\\
	0 \le &l(x_p) \le 4.5\\
	-0.1 \le &\theta(x_p) \le 0.1\\
	-1 \le &F_t(x_p) \le 1\\
	0 \le &F_h(x_p) \le 8.
\end{aligned}
\end{equation}
Additionally, a first-order polynomial degree was used for the collocation method in the solver, the problem was solved in $100$ control intervals, and the trolley and payload masses were respectively set to $m_1=1.2kg$ and $m_2=0.6kg$.

Finally, the position of each stack was assigned as in (\ref{eq:position}), where $c \in \mathbb{R}^{m}$ represents the positions where the stacks are along the $i\text{-axis}$. In this example, $9$ stack positions were assigned, separated by $10cm$ from each other.
\begin{equation}
	\label{eq:position}
		c = 
	\begin{bmatrix}
	0.1 & 0.2 & 0.3 & 0.4 & 0.5 & 0.6 & 0.7 & 0.8 & 0.9
	\end{bmatrix}.
\end{equation} 
Moreover, the height of each stack was represented as in (\ref{eq:height})
\begin{equation}
\label{eq:height}
	\begin{aligned}
	\rho = 
	\begin{bmatrix}
	0.5 & 1.0 & 1.0 & 1.0 & 2.0 & 2.0 & 2.4 & 2.5 & 1.0
	\end{bmatrix},
	\end{aligned}
\end{equation}
where $\rho \in \mathbb{R}^{m}$ represents the height of each of the stacks, corresponding to the assigned position in (\ref{eq:position}). Also, the width of the stacks was considered, here $0.08m$. From these, the function values in the bound constraints (\ref{eq:height_constraint_space}) can be constructed.

\subsection{Results}

\begin{figure}[h]
	\centering
	\includegraphics[scale=0.18]{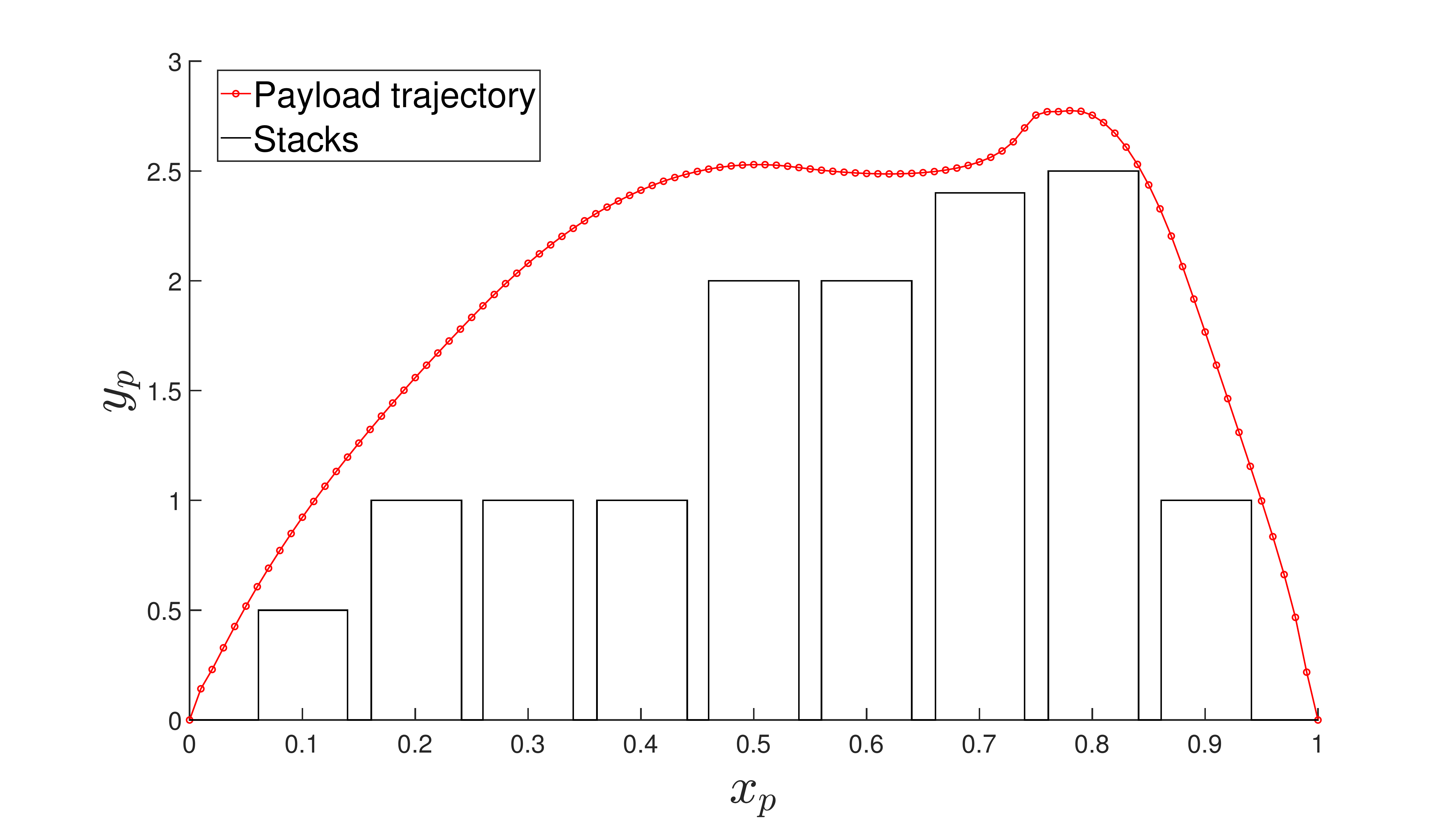}
	\caption{The path followed by the payload subject to stack constraints.}
	\label{fig:ex1-trajectory}
\end{figure}

\begin{figure}[h]
	\centering
	\includegraphics[scale=0.24]{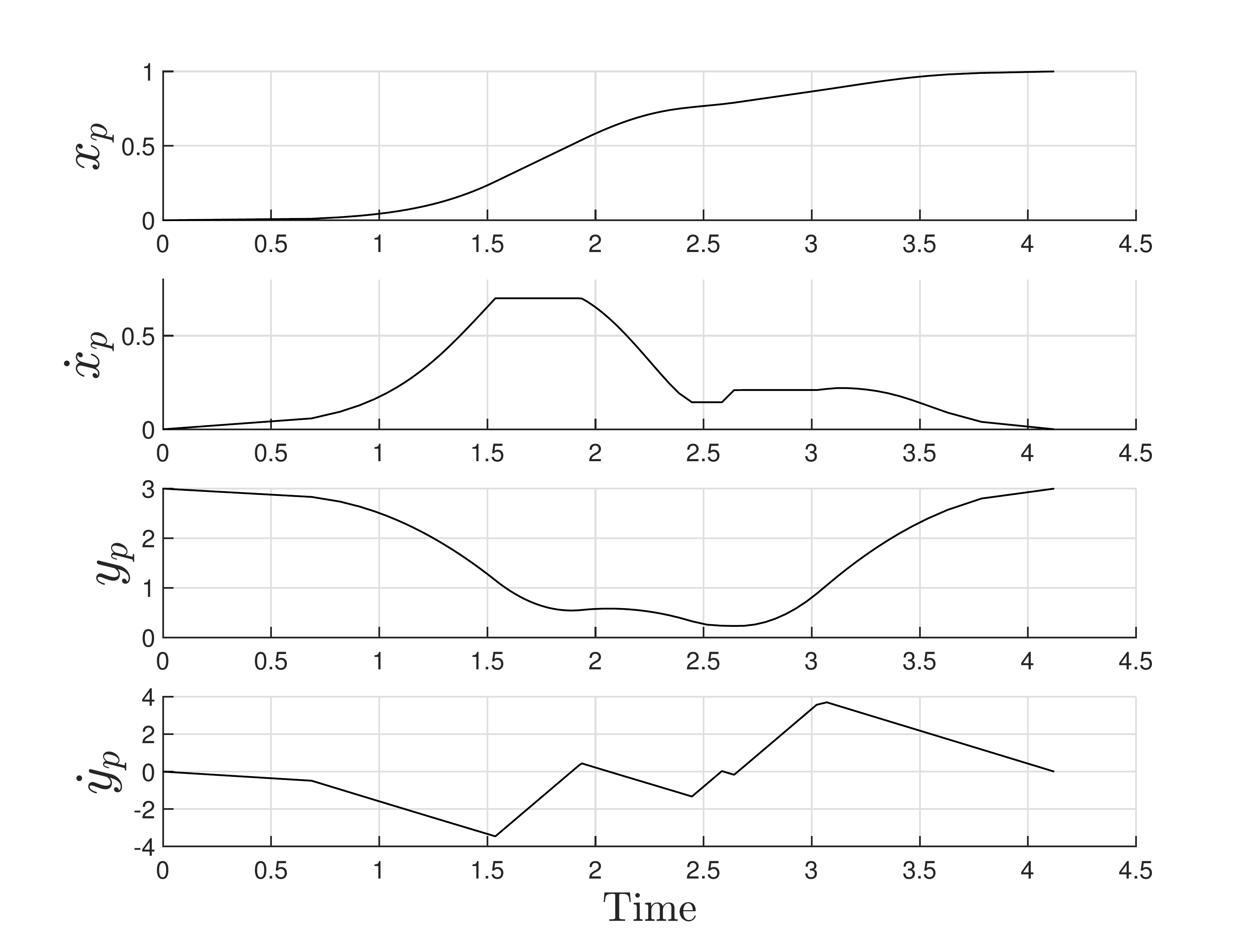}
	\caption{The payload spatial-coordinates trajectories.}
	\label{fig:ex1-states1}
\end{figure}

\begin{figure}[h]
	\centering
	\includegraphics[scale=0.24]{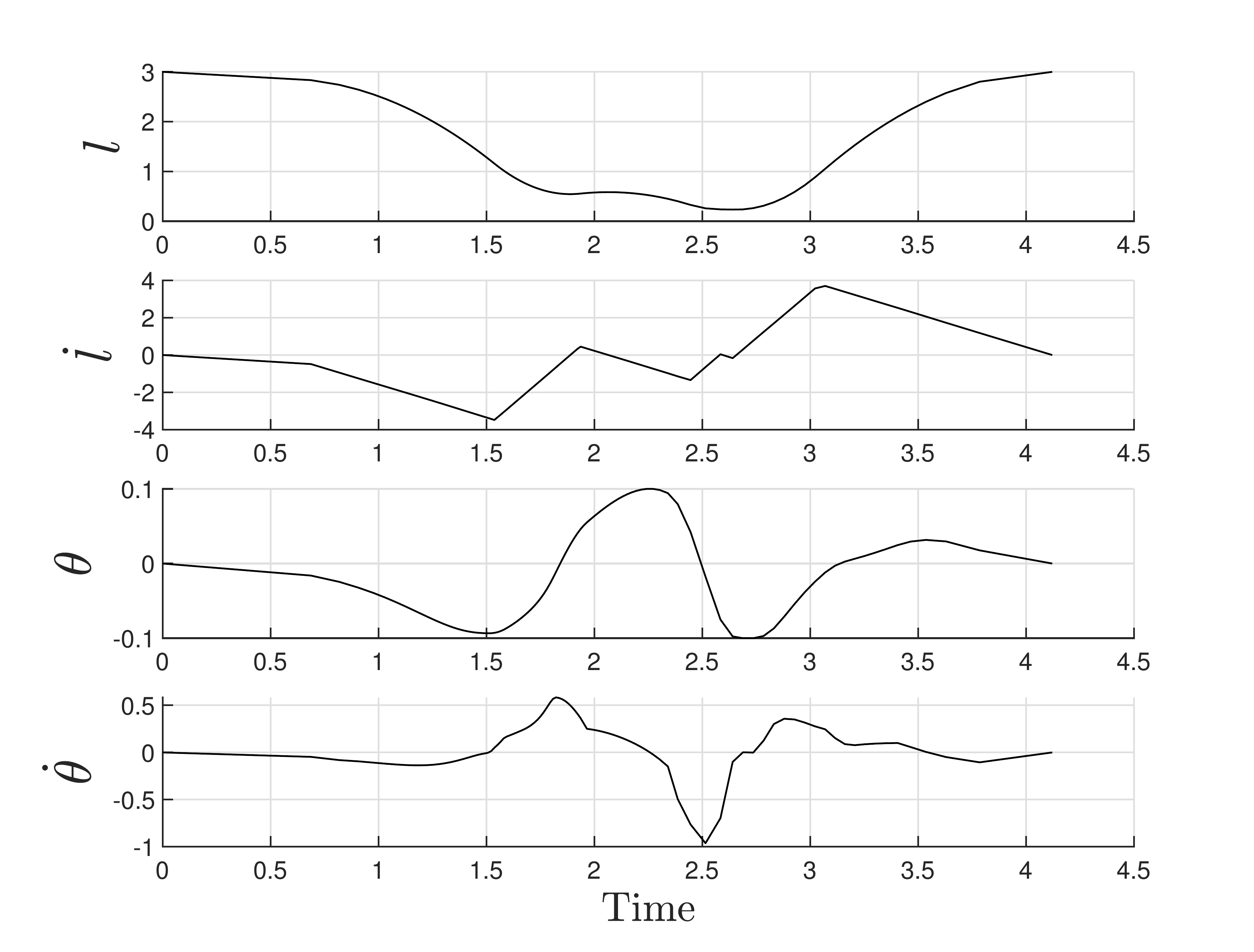}
	\caption{The payload hoisting and sway trajectories.}
	\label{fig:ex1-states2}
\end{figure}

\begin{figure}[h]
	\centering
	\includegraphics[scale=0.24]{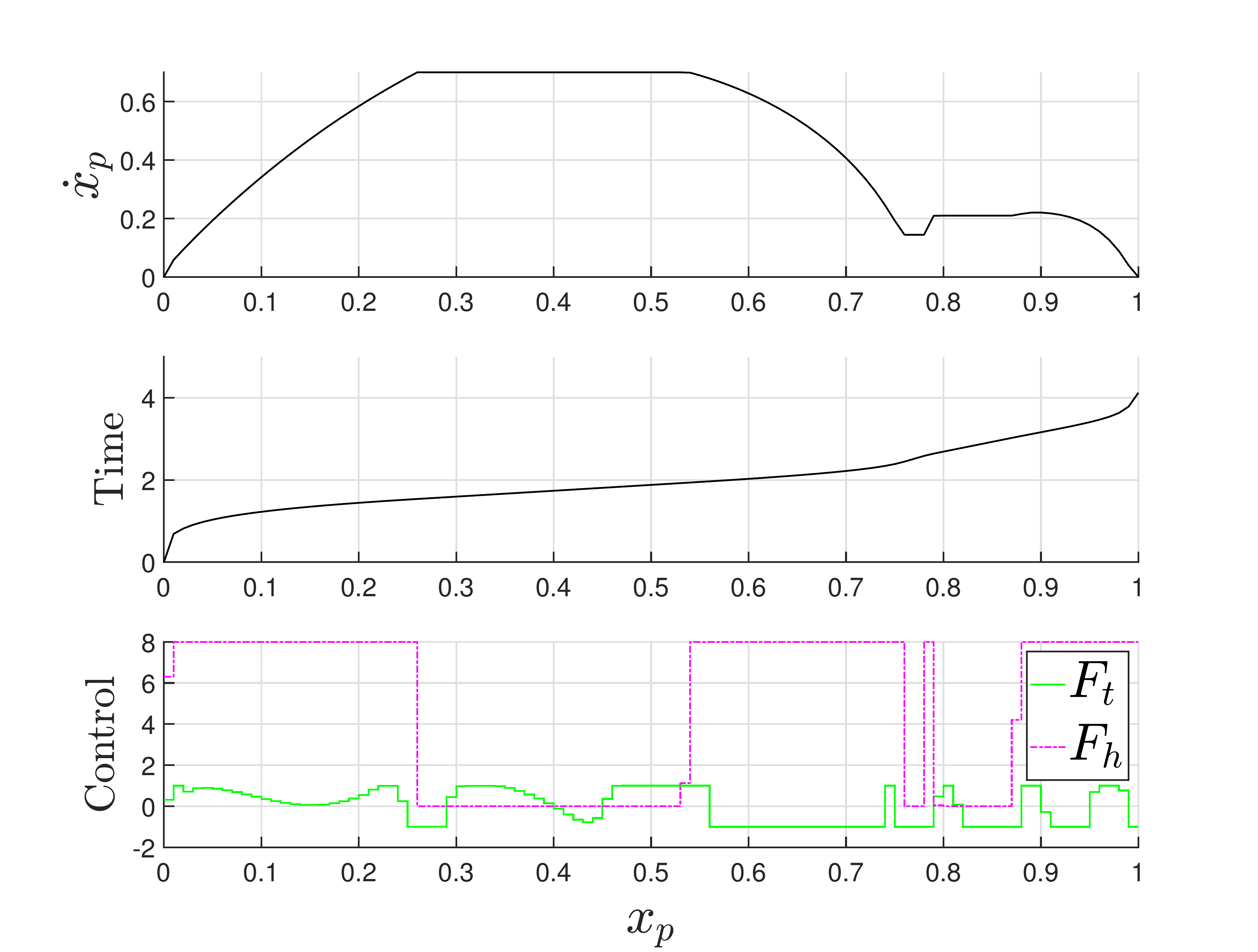}
	\caption{How $\dot{x}_{p}$, time and control inputs evolve with the spatial coordinate.}
	\label{fig:ex1-extra}
\end{figure}

\autoref{fig:ex1-trajectory} shows the path followed by the payload, given the stack heights (\ref{eq:height}) at the positions (\ref{eq:position}).
Figures \ref{fig:ex1-states1} and \ref{fig:ex1-states2} show how the payload positions, hoisting and sway evolve with time. Note that though they correspond to the decision variables in (\ref{eq:original-state-eq}), they are the result of problem solved as in (\ref{eq:opt-reformulated}).
Lastly, \autoref{fig:ex1-extra} shows the velocity $\dot{x}_p$, time (which now is a decision variable), and control inputs $F_{t}$ and $F_{h}$ evolving along $x_{p}$.

\subsection{Discussion} \label{sec:discussion}

The presented numerical example shows the simulation results for the optimization control problem, solved off-line after a change of variable. The aim was to illustrate that with the variable change, the description of the stack heights becomes trivial and easier to represent numerically. This is done by defining their positions and corresponding heights in $c$ and $\rho$, respectively. These are then implicitly defining a function $s(x_p)$, which addresses the stacks' height in each discretization interval. Moreover, Figs. \ref{fig:ex1-trajectory} to \ref{fig:ex1-extra} show that the payload was transferred from the initial to the final position avoiding the stacks and not violating the predefined constraints. Hence, we believe that the results have further strengthened our idea that this problem becomes trivial with the discretization of $x_p$.

We are aware that our work still has some limitations. As mentioned before, a natural one is that the solution enforces no sway condition since the payload is moving monotonically in one direction. Furthermore, the distance traveled by the payload in each discretization interval varies along the loading and unloading process. For instance, as seen in \autoref{fig:ex1-trajectory}, they were particularly long in the end, after a tall stack. Thus, since the problem is uniformly discretized, most of the dynamics that are not captured might be in those parts. This can be easily addressed by non-uniform discretization. Notice that these regions, where more time is spent, are strongly influenced by the stacks' disposition. 

Perhaps the biggest limitation of this work is that, though the objective function and the constraints on the container heights are now convex, the dynamics of the system remain non-convex. This might lead to a solution at a local minimum. However, it is important to note that circumventing these limitations is not in the scope of this work. Moreover, non-convexity is simply a consequence of the high-fidelity model of the system. Additionally, the problem may be extended to 3D, however, we believe that this is not necessary, given the nature of ship-to-shore crane operations.

\section{CONCLUSIONS AND FUTURE WORK} \label{sec:conclusion}

We have proposed a new approach to deal with the different heights that the container stacks can assume during the process of loading and unloading container ships. With this in mind, a variable change has been performed and the representation of the stack heights became trivial in the optimization problem. Subsequently, a small-scale example has led us to confirm it. Thus, this approach has the potential to be successfully applied to real crane operations and enhance their automation. 

Future work will focus on going beyond the point-mass assumption and incorporate more physical and geometric constraints to the setup. Moreover, we will investigate how to deal with it when the initial or final position is in between stacks. Additionally, different densities of discretization will be addressed.

\addtolength{\textheight}{-3cm}   
%


\bibliographystyle{IEEEtran}
\bibliography{bib/acc_2022}

\end{document}